\documentclass[review]{elsarticle}

\usepackage{lineno,hyperref}
\modulolinenumbers[5]

\journal{Journal of \LaTeX\ Templates}

\usepackage{graphicx}
\begin{document}

\begin{frontmatter}

\title{Predicting COVID-19 cases using Bidirectional LSTM on multivariate time series}


\author{Ahmed Ben Said}
\author{Abdelkarim Erradi}
\author{Hussein Ahmed Aly}
\author{Abdelmonem Mohamed}
\address{Computer Science \& Engineering Department, College of Engineering, 2713, Doha, Qatar}
\address{\{abensaid, erradi, ha1601589, am1604044\}@qu.edu.qa}


\begin{abstract}
\begin{itemize}
    \item \textbf{Background}: To assist policy makers in taking adequate decisions to stop the spread of COVID-19 pandemic, accurate forecasting of the disease propagation is of paramount importance.
    \item \textbf{Materials and Methods}: This paper presents a deep learning approach to forecast the cumulative number of COVID-19 cases using Bidirectional Long Short-Term Memory (Bi-LSTM) network applied to multivariate time series. Unlike other forecasting techniques, our proposed approach first groups the countries having similar demographic and socioeconomic aspects and health sector indicators using K-Means clustering algorithm. The cumulative cases data for each clustered countries enriched with data related to the lockdown measures are fed to the Bidirectional LSTM to train the forecasting model.
    \item \textbf{Results}: We validate the effectiveness of the proposed approach by studying the disease outbreak in Qatar. Quantitative evaluation, using multiple evaluation metrics, shows that the proposed technique outperforms state-of-art forecasting approaches.
    \item \textbf{Conclusion}: Using data of multiple countries in addition to lockdown measures improve accuracy of the forecast of daily cumulative COVID-19 cases. 
\end{itemize}

\end{abstract}


\end{frontmatter}


\section*{Introduction}
In December 2019, Wuhan, the capital of Central China's Hubei province, with 11 millions population, has witnessed the outbreak of a new coronavirus (COVID-19) \cite{yang,review}. The virus has propagated in China then all over the world. On the 11\textsuperscript{th} March 2020, with more than 280k cases and more than 4000 deaths worldwide, it has been declared as a global pandemic by the World Health Organization (WHO) \cite{who}. Within few months, the number of cases has exponentially grown to more than 17 millions and more than 670k deaths by end of July 2020. \newline
Various research works have been conducted to get better insights about the propagation of the virus and the evolution of number of cases and deaths while racing against the time to develop an effective vaccine. These works can be categorized into machine learning based and mathematical modeling-based.\newline
Saba et al. \cite{egypt} studied the propagation of COVID-19 in Egypt and applied a nonlinear autoregressive artificial neural network to forecast the virus prevalence. The authors modeled the confirmed cases as time series and compared their approach against Auto Regressive Integrated Moving Average (ARIMA) model. Both approaches are used to forecast the cumulative COVID-19 cases for ten days (1 to 10 April 2020) using the confirmed cases reported on March.
In \cite{round}, the authors applied exponential smoothing approach and conducted five rounds of forecast of cumulative confirmed cases globally starting from the 1\textsuperscript{th} February till 21\textsuperscript{th} March 2020. The authors emphasized that forecasts related to the virus outbreak must be an integral part of any decision-making process particularly in high risk cases. Indeed, this enables authorities to explore various 'what if' scenarios in order to assess the implication of any decision.
Ahmar et al. \cite{suttearima} proposed to apply a variety of ARIMA, called SutteARIMA, for  short-term forecast of COVID-19 cases in Spain and the impact on the Spanish Market Index (IBEX). Data from February 12 till April 2 are used to train the model to forecast the data from April 3 to 9. The Mean Absolute Percentage Error (MAPE) metric is calculated to assess the fitting accuracy. The findings showed that SutteARIMA outperformed ARIMA model. 
In \cite{brazil}, the authors compared six prediction techniques to forecast the cumulative cases in ten Brazilian states: ARIMA, cubist regression, random forest, ridge regression, support vector regression, and stacking-ensemble learning. The prediction is conducted for multiple time horizons: one day, three days and six days ahead. Chimula et. \cite{canada} studied the propagation of the virus in Canada using Long Short-Term Memory (LSTM) neural network, known to be efficient with sequential data. The results show that Canada had a linear growth of number of cases until March 16, 2020 followed by an exponential growth. It has been estimated that the ending point of the outbreak is around June. Maleki et al. \cite{TP_SMN_AR} applied TP-SMN-AR, a variation of autoregressive models to forecast the confirmed and recovered number of cases worldwide. This prediction is conducted for the period between April 21 till April 30.
\newline 
Mathematical models of infectious disease have been also applied in attempt to obtain better insight about the virus outbreak. Kuniya \cite{japan} applied the SEIR model to predict the epidemic peak in Japan from 15 January to 29 February. SEIR provides a mathematical formulation to describe the transmission of a disease from an individual to another. These individuals pass through four states: susceptible (S), exposed (E), infectious (I) and recover (R). The study showed that the basic reproduction number $R_0$- "the average number of secondary infections produced by a typical case of an infection in a population where everyone is susceptible" \cite{ro}- is 2.6 with a 95\% Confidence Interval 2.4-2.8. The SEIR model also showed that the peak will occur on early-middle summer 2020. Furthermore, some epidemiological conclusions are drawn: the intervention has great implication on delaying the epidemic peak. It also must be conducted over a long period to ensure effective reduction of the epidemic size. In \cite{algeria}, the authors applied SIR model to predict the daily cases in Algeria. SIR takes into account the number of susceptible cases (S), the number of infected cases (I) and the number of Recovered cases (R). The model showed that the peak is expected on July 24 at worst and that the disease will disappear between September and November. Roosa et al. \cite{china_3_models} used three phenomenological models: the generalized logistic growth model \cite{glm}, the Richards model \cite{richards} and a sub-epidemic wave model \cite{wave} for real-time forecasts of the COVID-19 cumulative number of confirmed reported cases in Hubei province, China. These models were previously applied to forecast several infectious diseases including Ebola, SARS, pandemic Influenza, and Dengue. Authors in \cite{weather} studied the effect of weather on the spread of COVID-19. Using the daily cases in 50 US states between January 1 and April 9 in addition to temperature and absolute humidity information, the authors identified the vulnerable narrow absolute humidity range. States with absolute humidity between 4 and 6 g/m$^3$ have significant spread with more than ten thousands cases. The findings are used to determine the Indian regions with potential vulnerability to weather based spread. \newline
It is widely known that lockdown measures, e.g. restriction on gathering, school and workplace closing, public transport shutdown and international travel controls, are needed for  halting the spread of the virus. Atalan et al \cite{lockdown1} conducted a data analysis and showed evidence that lockdown can contribute in suppressing COVID-19 pandemic. Dawoud \cite{lockdown3} emphasized on the importance of preventive measures including social distancing and mask usage for efficient lockdown exit strategy. Sahoo et al \cite{lockdown4} conducted a data-driven approach to analyze the effect of lockdown in India. The authors showed that after six weeks of lockdown, the infection rate reached three times lower compared to the initial one. Hence, the lockdown measures are quintessential to manage such pandemic. However, these measures are rarely considered when forecasting COVID-19 daily or cumulative cases. Furthermore, most COVID forecast methods typically rely on limited data of a single country. Yet  countries having common demographic and socio-economic properties and similar health sector indicators can exhibit similar pandemic patterns. Our contribution consists of first grouping countries having similar demographic and socio-economic properties and health sector indicators then using COVID-19 data from each cluster to build the prediction model. This yields a richer dataset for training. Furthermore, we propose a deep learning based forecasting approach using a Bidirectional LSTM (Bi-LSTM). This type of neural network not only relies on the past data to predict the future, but it also enables learning from the future to predict the past. By adopting such learning framework, Bi-LSTM provides better understanding of the learning context \cite{bilstm_paper}. Additionally, to train our Bi-LSTM-based model, we use multivariate time series consisting of the daily cumulative number of cases and time series describing the lockdown measures: the school closing, workspace closing, restriction on gathering, public transport closing and international travel controls.
The proposed Bi-LSTM on multivariate time series allows multiple dependent time series to be modelled together to account for the correlations cross and within the series capturing variables changing simultaneously over time.
\section*{Materials and Methods}
\begin{figure}[!h]
	\centering
	\includegraphics[scale=.66]{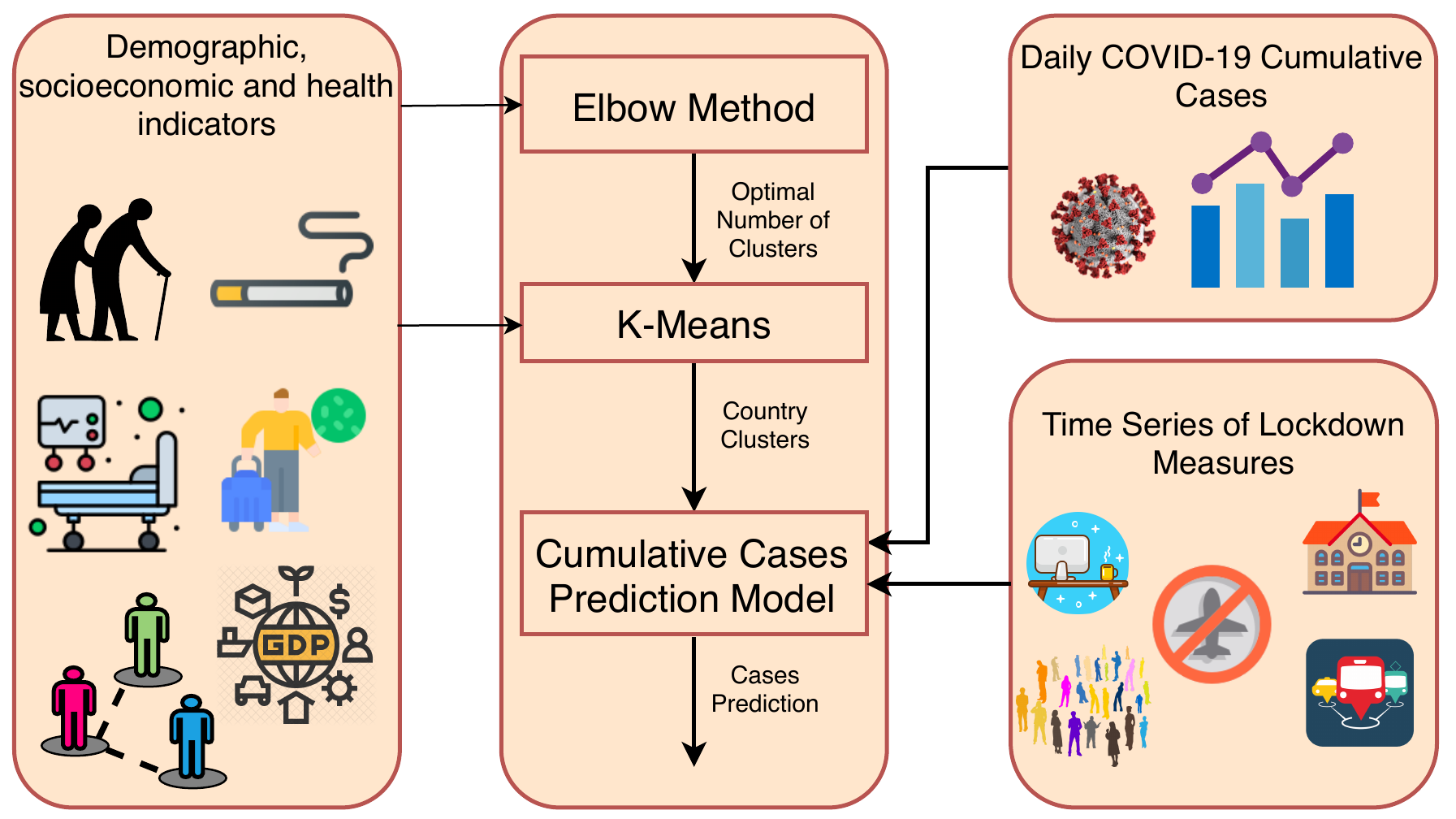}
	\caption{Overview of the proposed prediction approach of daily cumulative cases of COVID-19 using Bi-LSTM on multivariate time series}
	\label{approach}
\end{figure}

We depict in Fig. \ref{approach} the overall approach to predict the daily cumulative cases of COVID-19. First, we collect data describing the demographic and socioeconomic properties and health sector of countries of the world. These data are clustered to identify group countries that have similar properties. We first apply the Elbow method to determine the optimal number of clusters to pass as an input parameter for the K-Means algorithm. Next, given a particular country, we identify its cluster. Multivariate time series are then constructed consisting of daily cumulative cases of all countries belonging to the cluster in addition to time series describing the level of lockdown measures associated to travel control (border closing), school closing, workplace closing, public transport shutdown and public gathering ban. The multivariate time series is used to train a deep learning Bi-LSTM network to forecast future cumulative number of cases. It is worthwhile to mention that this approach is applicable for any country to forecast its daily cumulative COVID-19 cases.
\subsection*{\textbf{Clustering countries based on demographic, socioeconomic and health sector indicators}}
We describe in this section the demographic, socioeconomic and health sector indicators used to cluster countries. Then we present the approach used to cluster countries having similar properties. This yields a richer dataset for training COVID-19 cumulative cases prediction model per countries cluster. 

\subsubsection*{Demographic, socioeconomic and health sector indicators data}
These data have been collected from the Department of Economic and Social Affairs of the United Nations and the Organisation for Economic Cooperation and Development. The data include:
\begin{itemize}
    \item Median age per country.
    \item Population percentage of age groups per 4 years interval e.g. 4-9 year, 10-14 years etc.
    \item Country population and density.
    \item The percentage of urban population.
    \item Gross Domestic Product (GDP) per capita.
    \item The number of hospitals per 1000 people.
    \item Death rate from lung diseases per 100k people for female and male.
\end{itemize}

\subsubsection*{Countries clustering}
To discover countries having similar characteristics we applied K-Means clustering algorithm \cite{jain,abensaid} to identify similar members among the data points.
Let $X=\{x_1,x_2,...,x_n\}$ be the set of d-dimensional points we seek to cluster into $K$ clusters. In other words, we attempt to assign each $x_i$, $i=1,...,n$ to a cluster $c_k$, $k=1,...,K$. K-Means partitions the data such that the squared error between the mean of a cluster and the data points, members of the clusters, is as low as possible. Let $m_k$ be the mean of cluster $c_k$. The squared error between a cluster center and its members is defined as:
\begin{equation}
    J(c_k) = \sum\limits_{x_i \in c_k} || x_i - m_k||^2 
\end{equation}
K-Means seeks to minimize the sum of the squared errors:
\begin{equation}
    J(C) = \sum\limits_{k=1}^K \sum\limits_{x_i \in c_k} || x_i - m_k||^2
    \label{kmeans}
\end{equation}
Where $C$ is the set of clusters. To minimize Eq. \ref{kmeans}, the following algorithm is applied:
\begin{enumerate}
    \item Randomly assign $K$ cluster centers and repeat step 2 and 3.
    \item Assign each data point to the closest cluster center.
    \item Calculate the new cluster centers.
\end{enumerate}
However, K-Means requires the number of clusters to be known. Hence, we applied the Elbow method to determine the optimal number of clusters for which the obtained partition is compact, i.e. low $J(C)$. Naturally, adding more clusters would result in even more compact partition which may lead to over-fitting. Hence, the variation of $J(C)$ with respect to $K$ would exhibit first a sharp decrease followed by a slow one. The Elbow method recommends to select the number of cluster corresponding to the elbow of the curve $J(C)$ vs $K$.

\subsection*{\textbf{Bi-LSTM for COVID-19 cumulative cases prediction}}
After K-Means is applied to group countries, we collect the multivariate time series data for the countries of each cluster to train a prediction model using a Bi-LSTM deep neural network. The motivation is to strengthen the prediction accuracy by forcing the network to train not only on past data to predict the future but also to train it on the future data to predict the past.

\subsubsection*{Multivariate time series data}
The multivariate time series has more than one time-dependent variable. Intrinsically, these variables are also dependent on each others. Indeed, it is confirmed that lockdown measures have significant impact on the evolution of the cumulative number of COVID-19 cases. Our times series consists of:
\begin{itemize}
    \item Cumulative COVID-19 cases per day. These data are widely available and several APIs provided by government agencies can be queried for this information. We collect data from February 15th to July 31st.
    \item School closing: This time series describes the level of lockdown imposed on schools where 0 indicates no measures,1- recommend closing, 2- require closing (only some levels or categories, e.g. just high school, or just public schools) and
    3- require closing all levels.
    \item Workplace closing, where 0 indicates no measures,1- recommend closing (or recommend work from home), 2- require closing (or work from home) for some sectors or categories of workers and 3- require closing (or work from home) for all-but-essential workplaces (e.g. grocery stores, doctors).
    \item Restrictions on gatherings: where 0 indicates no restrictions, 1- restrictions on very large gatherings (the limit is above 1000 people), 2- restrictions on gatherings between 101-1000 people, 3- restrictions on gatherings between 11-100 people, 4- restrictions on gatherings of 10 people or less.
    \item Public transport shutdown where 0 indicates no measures, 1- recommend closing (or significantly reduce volume/route/means of transport available) and 2- require closing (or prohibit most citizens from using it)
    \item International travel controls where 0 indicates no restrictions,1- screening arrivals, 2- quarantine arrivals from some or all regions,
3- ban arrivals from some regions and 4- ban on all regions or total border closure.
\end{itemize}

\subsubsection*{Training the prediction model for COVID-19 cumulative cases}
\begin{figure}[h!]
	\centering
	\includegraphics[scale=.7]{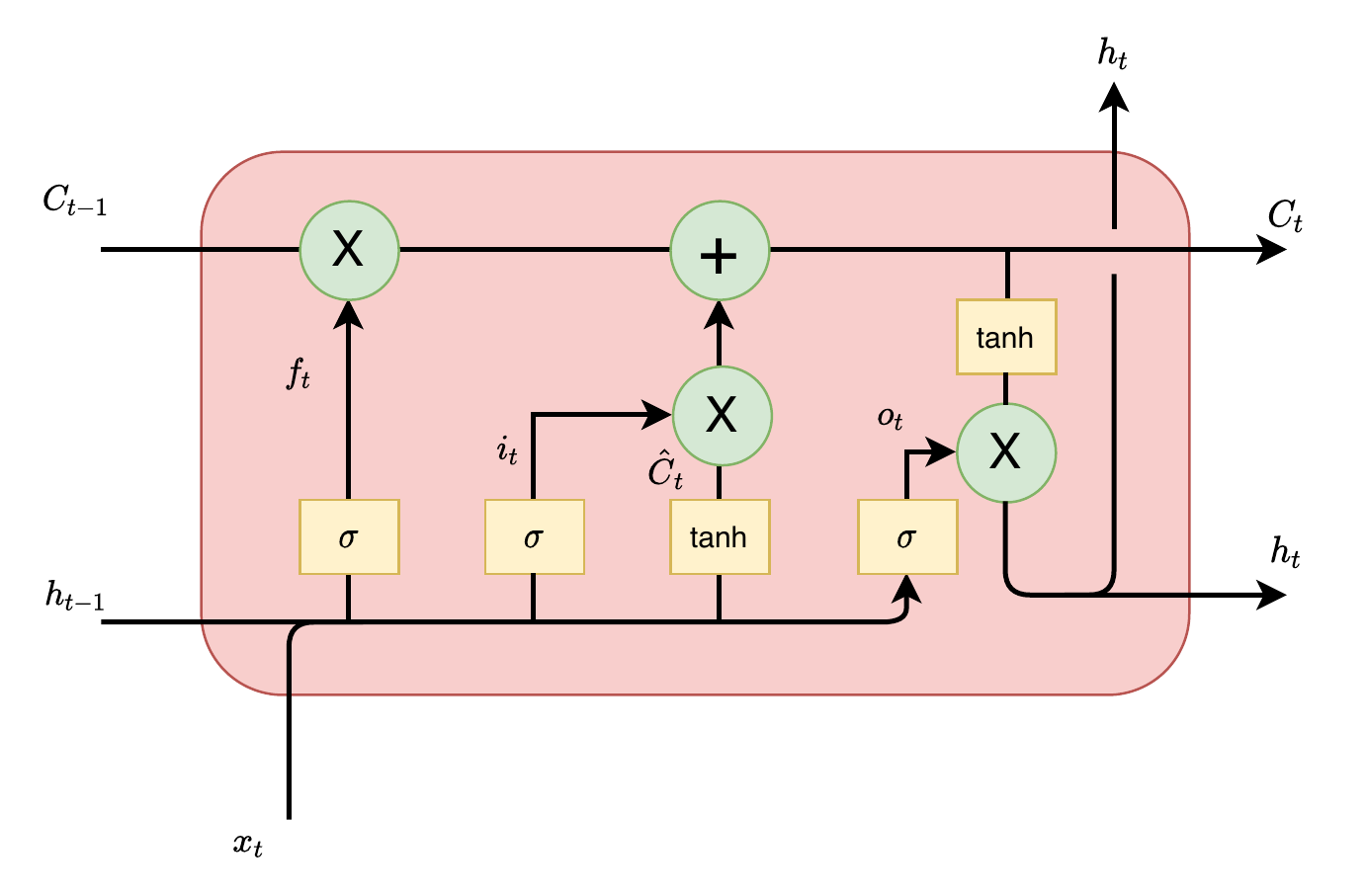}
	\caption{Long Short-Term Memory (LSTM) cell}
	\label{lstm_cell}
\end{figure}

\begin{figure}[h!]
	\centering
	\includegraphics[scale=.66]{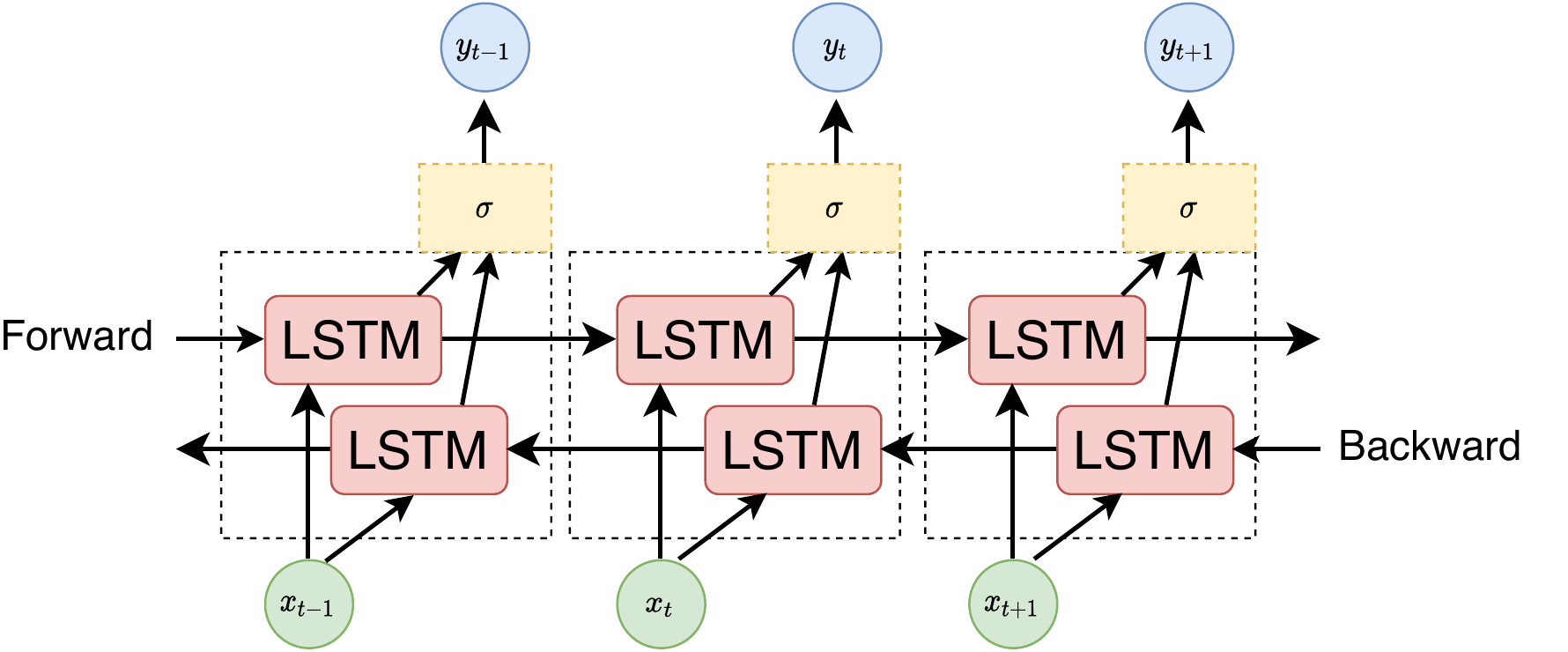}
	\caption{Unfolded architecture of Bidirectional LSTM}
	\label{bi-lstm}
\end{figure}

The building block of the network is the LSTM cell depicted in Fig. \ref{lstm_cell}.
Given the current value $x_t$, the previous hidden state $h_{t-1}$ and the previous state $C_{t-1}$, the following transformations are applied:
\begin{equation}
    f_t = \sigma\Big(W_f\cdot [h_{t-1}, x_t] + b_f \Big)
\end{equation}
\begin{equation}
    i_t=\sigma\Big(W_i[h_{t-1},x_t] + b_i\Big)
\end{equation}
\begin{equation}
    \hat{C}_t=tanh\Big(W_C[h_{t-1},x_t] +b_c\Big)
\end{equation}
\begin{equation}
    C_t=f_t*C_{t-1}+i_t*\hat{C}_t
\end{equation}
\begin{equation}
    o_t = \sigma\Big( W_o[h_{t-1},x_t]+b_o\Big)
\end{equation}
\begin{equation}
    h_t = o_t*tanh(C_t)
\end{equation}
Where $\sigma$ and $tanh$ are the sigmoid and hyperbolic tangent function respectively. $f_t$ is the forget gate, $i_t$ is the input gate and $o_t$ is the output gate. $W$ and $b$ are the weight matrix and bias vector respectively. $[\cdot]$ is the concatenation operator and $*$ is the dot product.
Hence, an LSTM layer consists of a sequence of LSTM cells and the sequence data are fed in a forward way. Bi-LSTM includes another LSTM layer for which the data is fed in backward way as depicted in Fig. \ref{bi-lstm}. By stacking multiple Bi-LSTM layers, i.e. the output of one layer is fed to another one, a deep neural network can be trained to forecast the next day cumulative number of cases. The network is trained using backpropagation \cite{backprop} \cite{backprop2} algorithm to minimize the mean squared error between the actual daily cumulative cases and the value predicted by the network. 
\section*{Experiments}
The proposed technique is versatile. Indeed, the forecast can be applied for any country. In our experiment, We aim at using information from the previous 6 days to predict the next day cumulative cases. We focus on Qatar as a use-case and we analyze and assess the forecast performance of the proposed technique and compare against multiple techniques with multiple scenarios.
\subsection*{\textbf{Evaluation approach}}
We analyze the performance by:
\begin{itemize}
    \item Comparing the prediction performance of the proposed approach against LSTM model. We show the benefit of: 1- Training the learning models on data from all countries in the same cluster. 2- Including lockdown information in the training data. 
    \item  Comparing against state-of-art techniques including ARIMA, Simple Moving Average with 6-day window and Double Exponential Moving Average.
    \item Evaluating the prediction accuracy by reporting the Root Mean Square Error (RMSE), Mean Absolute Error (MAE), Coefficient of Residual Mass (CRM) and the Determination Coefficient R$^2$ where:
    \end{itemize}
    \begin{equation}
        RMSE = \sqrt{\frac{1}{n} \sum\limits_{i=1}^n (x_i -y_i)^2}
    \end{equation}
    \begin{equation}
        MAE = \frac{1}{n} \sum\limits_{i=1}^n \left| \frac{x_i-y_i}{x_i} \right|
    \end{equation}
    \begin{equation}
        CRM = \frac{\sum\limits_{i=1}^n y_i - \sum\limits_{i=1}^n x_i}{\sum\limits_{i=1}^n y_i}
    \end{equation}
    \begin{equation}
        R^2 = \frac{\Big(\sum\limits_{i=1}^n(x_i-\hat{x})(y_i-\hat{y})  \Big)^2}{\sum\limits_{i=1}^n (x_i-\hat{x})^2  \sum\limits_{i=1}^n (y_i-\hat{y})^2}        
    \end{equation}
Where $x_i$, $y_i$, $\hat{x}$ and $\hat{y}$ are the actual reported cumulative cases, predicted cumulative cases, average reported cumulative cases and average predicted cumulative cases respectively. The best prediction is the one achieving the lowest RMSE, MAE, the highest R$^2$ and the closest CRM value to zero.
\subsection*{\textbf{COVID-19 in Qatar}}
\begin{figure}[h!]
	\centering
	\includegraphics[scale=.4]{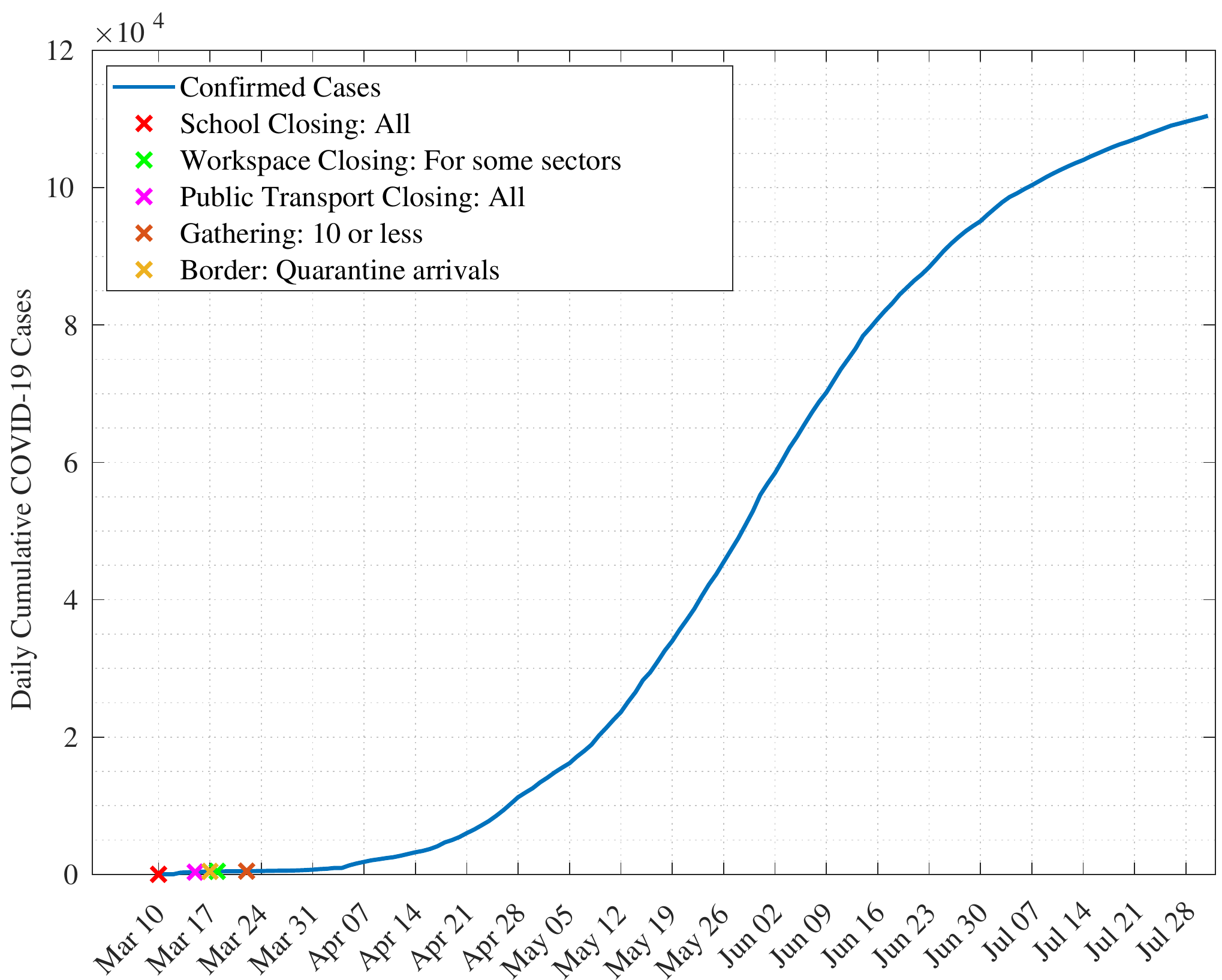}
	\caption{Cumulative COVID-19 cases in Qatar with lockdown measures}
	\label{qatar}
\end{figure}

We illustrate in Fig. \ref{qatar} the evolution of the daily cumulative cases in Qatar from March 10 to July 31. Till end of March, the cumulative cases evolved in a linear trend. Then, numbers have started to grow exponentially till mid June. By mid June, the growth of number of cases has started to slow down. The first confirmed case has been reported on February 29. By July 31, 235 cases have been reported. All lockdown measures have been imposed on March. School were all closed on March 10. Then, all public transport services have been shutdown on March 15. Borders have been closed on March 17 and quarantine is required on arrivals from all regions for nationals. Workplace have been also closed for some sectors on March 18 and public gathering for more than 10 persons has been prohibited on March 22. By July 31, the total cumulative cases reached 110460. \newline
\subsection*{\textbf{Data clustering}}
\begin{figure}[!h]
	\centering
	\includegraphics[scale=.35]{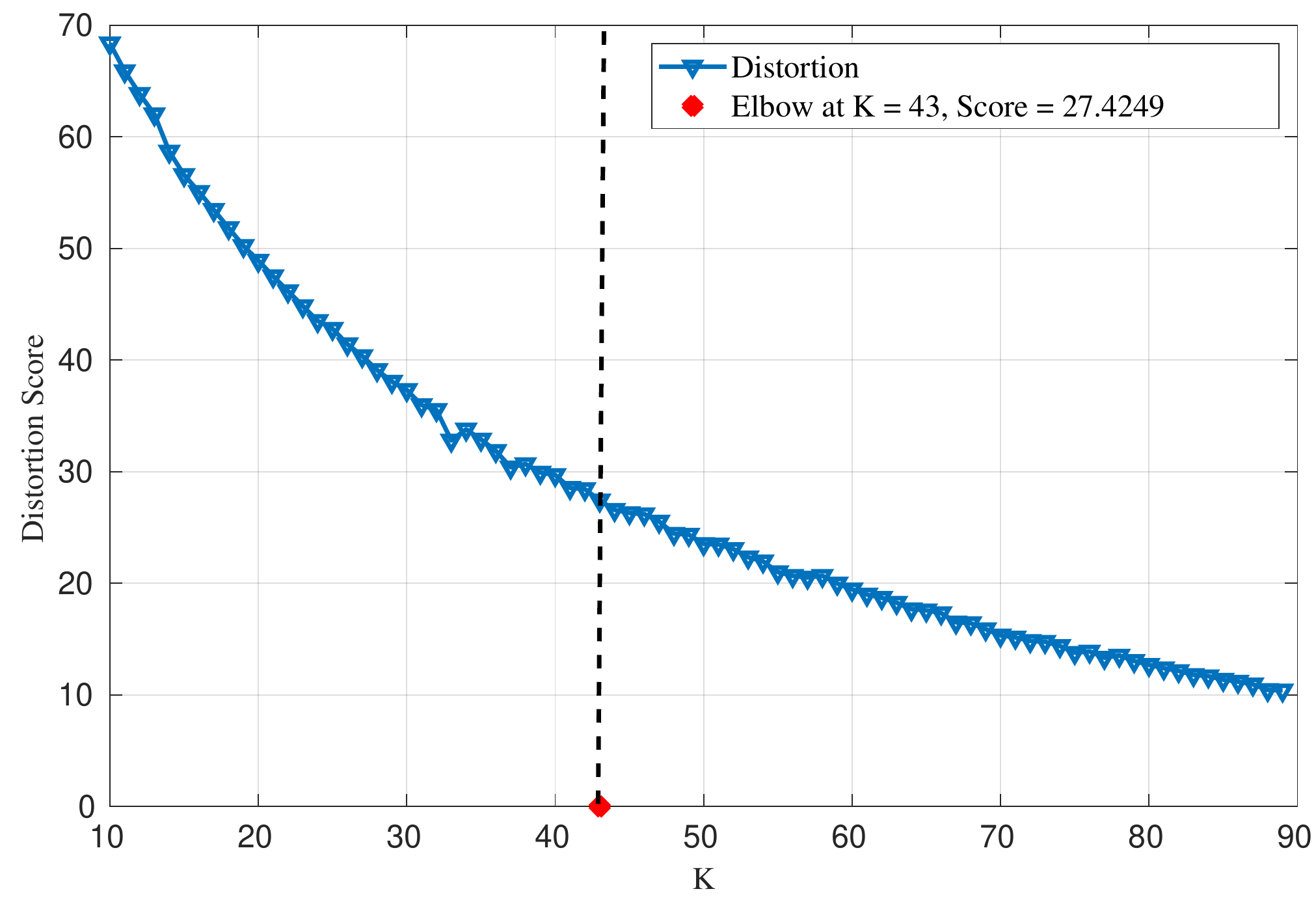}
	\caption{Distortion score for different numbers of clusters. Elbow corresponds to $K=43$}
	\label{elbow}
\end{figure}

\begin{figure}[h!]
	\centering
	\includegraphics[scale=.35]{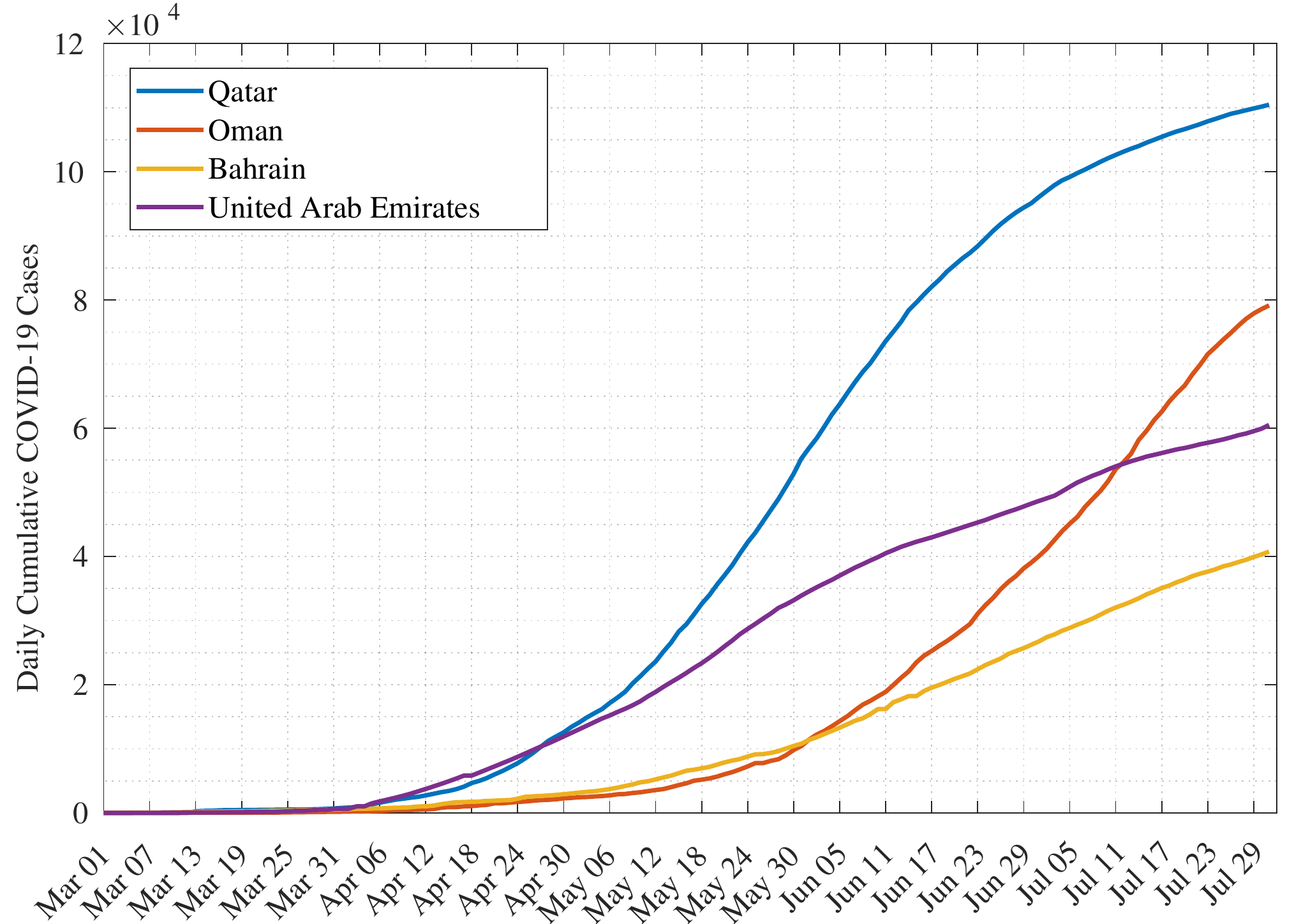}
	\caption{Cumulative COVID-19 cases of  countries having similar demographic and socioeconomic properties}
	\label{gcc}
\end{figure}

\begin{figure}[h!]
	\centering
	\includegraphics[scale=.4]{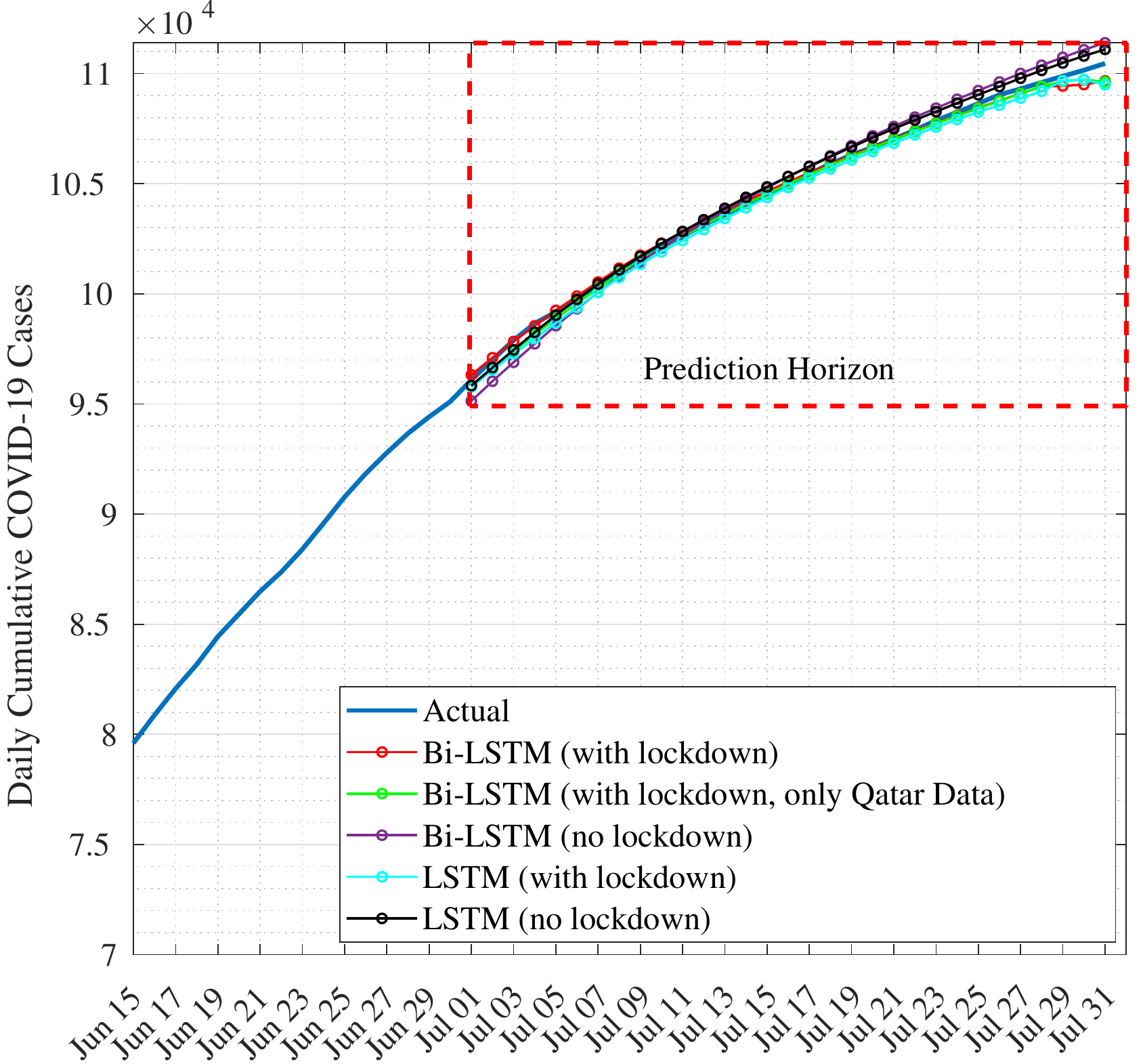}
	\caption{Forecasting results for Qatar using Bi-LSTM vs. LSTM models trained on Qatar cluster data}
	\label{dl_forecast}
\end{figure}

\begin{figure}[h!]
	\centering
	\includegraphics[scale=.3]{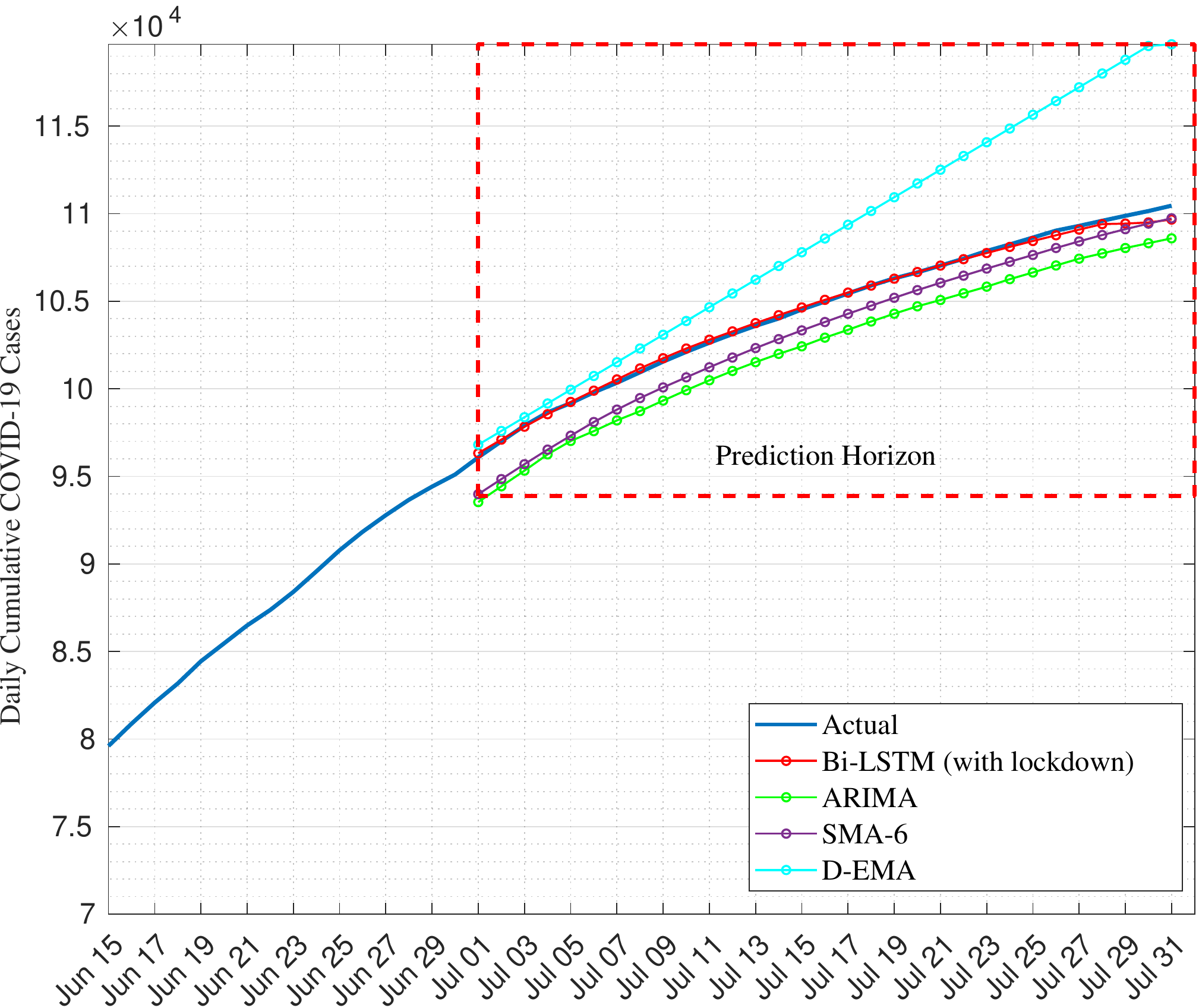}
	\caption{Forecasting results for Qatar using Bi-LSTM with lockdown compared to state-art time series forecasting approaches}
	\label{sota}
\end{figure}

By clustering the  socioeconomic and demographic properties data and the health sector indicators data, we intend to discover countries having similar properties. For the Elbow method, we use the distortion, i.e. the mean squared distances to the cluster centers, as a metric. Results are depicted in Fig. \ref{elbow}. The findings suggest that K = 43 corresponds to the Elbow and is the optimal number of clusters.
Clustering results using K-Means show that Qatar shares similar properties as Oman, Bahrain and United Arab Emirates (UAE).
Our findings also show that for example, Belgium, Canada, Finland, Sweden and United Kingdom are in the same group. \newline
Fig. \ref{gcc} shows the cumulative cases of Qatar, Oman, Bahrain and UAE. We notice that Qatar exhibits the most severe growth in number of cases, with exponential-like shape. UAE exhibits two linear trends. The first one, witnessed till end of March,  is linear with slow growth. Then, we notice a second linear trend with sharper growth in number of cases. For Bahrain, the increase is less severe compared to the two other countries. Oman initially had a linear increase of number of case. However, a sharper increase is witnessed starting from June.
We illustrate in Fig. \ref{dl_forecast} the actual growth of number of cases for Qatar and the forecasting results for LSTM and Bi-LSTM with and without lockdown information. Models are trained on data of all countries in the cluster.
For illustration purposes, we show data from June 15 to July 31. The findings show 
deep learning techniques succeeded in capturing the trend of cumulative cases in Qatar. The predictions curves are similar to the actual cumulative cases data. To further assess the prediction performance, we conduct quantitative analysis, detailed in Table. \ref{dl_table}. Results show that Bi-LSTM with lockdown information results in the lowest RMSE, MAE, the highest R$^2$ score and the closest CRM value to zero while the best performance is achieved when the model is trained on all data of the cluster to which Qatar belongs rather than Qatar data only. In fact, this is confirmed by both the RMSE and CRM values comparison. Results also allow us to confirm the importance of including lockdown information as they improved the performance of both LSTM and Bi-LSTM models.
We further compare the proposed approach against state-of-art time series forecasting approaches including ARIMA, Simple Moving Average with 6-day window (SMA-6) and Double Exponential Moving Average (D-EXP-EMA) . Fig. \ref{sota} illustrates the forecasting results. It clearly shows how the proposed technique outperformed other approaches. Infact, SMA-6 and ARIMA tend to under-estimate the total number of cases, while D-EMA over-estimates the number of cases. This performance is quantitatively confirmed by the evaluation metrics detailed in Table .\ref{results_sota}
\section*{Discussion}
Predicting cumulative COVID-19 cases is a challenging task as it depends on several complex and highly dependable parameters. The disease outbreak depends on the lockdown measures and how fast they are imposed. In our proposed approach, we aimed at incorporating several parameters to achieve accurate forecast by: 1- maximizing the data used to train a forecasting model by grouping countries having similar properties: 2- using a Bi-LSTM model trained on both  number of cases and lockdown measures. It has been confirmed that rushing towards easing lockdown measures has contributed in an increase of number of cases. This has been the case of Florida and Texas, USA. In fact, Florida reopened certain business on May 4 and Florida Keys businesses were allowed to reopen to visitors on June 1. In Texas, school districts were allowed to open. Both states witnessed significant growth in number of cases. The proposed solution may assist decision-makers to put short-term future plans to overcome the epidemic and to carefully choose the opening strategy.

\section*{Conclusion}
We proposed a deep learning based solution to forecast the daily cumulative COVID-19 cases. The proposed approach grouped countries having similar demographic socioeconomic and health sector properties in order to train the forecasting model on data associated to the cluster rather than data of each country separately. By using Bi-LSTM and including lockdown information in the forecasting data, the proposed approach achieved significant improvement in the prediction performance compared to state-of-art techniques with Qatar as a use case. In future, work, we will establish lockdown easing scenario and investigate the forecasting results in order to analyze the impact of the easing on the increase/decrease of number of cumulative cases.

\section*{Acknowledgment}
This work was made possible by COVID-19 Rapid Response Call (RRC) grant \# RRC-2-104 from the Qatar National Research Fund (a member of Qatar Foundation). The statements made herein are solely the responsibility of the authors.

\subsection*{\textbf{Cumulative cases prediction}}
\begin{table}[h!]
\label{dl_table}
\caption{Evaluation results of deep learning models}
\begin{tabular}{l|c|c|c|c|}
\cline{2-5}
\multicolumn{1}{l|}{} & RMSE & MAE & R$^2$ & CRM \\ \hline
\multicolumn{1}{|l|}{Bi-LSTM with lockdown} & 245.1 & 176.02 & 0.996 & -0.0003 \\ \hline
\multicolumn{1}{|l|}{Bi-LSTM with lockdown (only Qatar data)} & 258.24 & 175.22 & 0.996 & -0.0016 \\ \hline
\multicolumn{1}{|l|}{Bi-LSTM without lockdown} & 389.6 & 321.9 & 0.981 & -0.00065 \\ \hline
\multicolumn{1}{|l|}{LSTM with lockdown} & 373.03 & 325.6 & 0.99 & -0.00061  \\ \hline
\multicolumn{1}{|l|}{LSTM without lockdown} & 380.19 & 349.03 & 0.977 & 0.0071 \\ \hline
\end{tabular}
\label{results}
\end{table}

\begin{table}[h!]
\caption{Performance evaluation of Bi-LSTM with lockdown, ARIMA, SMA-6, and D-EXP-MA}
\begin{tabular}{c|c|c|c|c|}
\cline{2-5}
\multicolumn{1}{l|}{} & RMSE & MAE & R$^2$ & CRM \\ \hline
\multicolumn{1}{|c|}{Bi-LSTM with lockdown} & 245.1 & 176.02 & 0.996 & -0.0003\\ \hline
\multicolumn{1}{|c|}{ARIMA} & 2109.1 & 2099.84 & 0.744 & -0.02 \\ \hline
\multicolumn{1}{|c|}{SMA-6} & 1356.5 & 1287.4 & 0.89 & -0.012 \\ \hline
\multicolumn{1}{|c|}{D-EXP-MA} & 2110.2 & 1562.7 & 0.744 & 0.01 \\ \hline
\end{tabular}
\label{results_sota}
\end{table}
\end{document}